\documentstyle[12pt,epsfig]{article}
\begin{document}
\vspace{-2.0cm} 
\begin{flushright}
TIFR/TH/00-28 \\
hep-lat/0006015 \\
\end{flushright}
\bigskip
\begin{center}
\Large {\bf Suppressing monopoles and vortices : A possibly
smoother approach to scaling ?}  \\

\bigskip

\large {Rajiv V. Gavai\footnote {gavai@tifr.res.in}} \\
\bigskip
Department of Theoretical Physics \\
Tata Institute of Fundamental Research\\
Homi Bhabha Road \\
Mumbai 400005, India \\

\bigskip
\bigskip
{\bf ABSTRACT}\\ \end{center}
\bigskip
\noindent  
Suppressing monopoles and vortices by introducing large chemical
potentials for them in the Wilson action for the $SU(2)$ lattice
gauge theory, we study the nature of the deconfinement phase transition 
on $N_\sigma^3 \times N_\tau$ lattices for $N_\tau =$ 4,
5, 6 and 8 and $N_\sigma =$  8--16.  Using finite size scaling theory, we
obtain $\omega \equiv \gamma/\nu = 1.93 \pm 0.03$ for $N_\tau =$ 4,
in excellent agreement with universality. Corresponding
determinations for the $N_\tau =$ 5 and 6 lattices are also found to be
in very good agreement with this estimate.  The critical couplings for
$N_\tau$ = 4, 5, 6 and 8 lattices exhibit large shifts towards the strong
coupling region when compared with the usual Wilson action, and 
suggest a lot smoother approach to scaling.

\newpage

\begin{center}
\bf 1. INTRODUCTION \\
\end{center}
\bigskip

Quantum field theories need regularization schemes to control divergences.  
The regularization schemes, many different types of which have been used in 
performing calculations, do not themselves affect physics in any
manner, as they are eliminated at the end of all calculations.  Long
distance physics, such as confinement of quarks in quantum
chromodynamics or determination of the hadronic spectrum, is
conveniently studied using the lattice regularization.  There is a lot
of freedom in defining a lattice field theory.   In particular, a
variety of different choices of the lattice action correspond to the
same quantum field theory in the continuum.  While many numerical
simulations have been performed for the Wilson action\cite{Wil} for the
gauge theories, other choices, some motivated by the desire to find a
smoother continuum limit, have also been used.  These actions differ
merely by irrelevant terms in the parlance of the renormalization group:
in the naive classical continuum limit of $ a \to 0$, they all reduce to
the same Yang-Mills action and the differing terms are of higher order
in $a$.

Investigations of the deconfinement phase transition for mixed actions, 
obtained by extending the  Wilson action by addition  of
an adjoint coupling term, showed \cite{us1,us2,us3,Ste} surprising
challenges to the above notion of universality.  These 
actions\cite{BhaCre,CaHaSc} are

\begin{eqnarray}
S_{BC} = \sum_P \left( \beta \left(1 - {1\over2} {\rm Tr}_F U_P \right) + 
	 \beta_A \left(1 - {1\over3} {\rm Tr}_A U_P \right) \right)~,~
\label{sbc}
\end{eqnarray}

and

\begin{eqnarray}
S_V = \sum_P \left( {1 \over 2} \left(\beta  + \beta_V~\sigma_P\right) 
{\rm Tr}_F U_P \right) ~.~
\label{sv}
\end{eqnarray}

Here $U_P$ denotes the directed product of the basic link variables
which describe the gauge fields, $U_\mu(x)$, around an elementary
plaquette $P$. $F$ and $A$ denote that the traces are evaluated in the
fundamental and adjoint representations respectively and the formula
${\rm Tr}_A U = |{\rm Tr}_F U | ^2 -1$ is used.  $\sigma_P$ are $Z_2$ plaquette
fields associated with the plaquette P and the partition function in the
Villain case of eq.(\ref{sv}) has a sum over all possible values for 
each of them.  The first term in both the equations describes the
standard SU(2) Wilson action whereas the second term adds an adjoint
SO(3) action.  For zero adjoint coupling, i.e, for the Wilson action,
several finite temperature investigations have shown the presence of a
second order deconfinement phase transition. Its critical exponents have
been shown\cite{EngFin} to be in very good agreement with those of the
three dimensional Ising model, as conjectured by Svetitsky and 
Yaffe\cite{SveYaf}.  The verification of the universality conjecture
strengthened our analytical understanding of the deconfinement phase
transition which, however, came under a shadow of doubt by the results
for the mixed actions.  Following the deconfinement phase transition into
the extended coupling plane by simulating these actions at finite temperature, 
it was found on a range of temporal lattice sizes for both actions
(\ref{sbc}) and ({\ref{sv}) that:  

\begin{enumerate} 

\item[{a]}] The deconfinement transition was of second order, and in agreement 
with the conjectured universality exponents, for small values of the adjoint 
coupling.  It became definitely of first order for large values $\beta_A$
or $\beta_V$. 

\item[{b]}] The deconfinement order parameter acquired a nonzero value 
discontinuously at the transition point for large adjoint couplings.

\item[{c]}] There was no evidence of any other separate bulk transition at
those large adjoint couplings, as expected from the results of Refs. 
\cite{BhaCre,CaHaSc}.  In fact, simulations on larger symmetric
lattices even suggested\cite{self} a lack of a bulk phase transition
at that adjoint coupling where a first order deconfinement transition for a 
lattice of temporal size four was clearly seen.

\end{enumerate}

Recently it was shown\cite{GavMan,DatGav} that suppression of some lattice
artifacts such as $Z_2$ monopoles and vortices do restore the
universality for the action (\ref{sv}): no first order deconfinement
transition was found in the entire coupling plane in that case.  In this
paper, we address this question for the action (\ref{sbc}) in the same
manner and find that unlike the Villain action, one gains an additional bonus.  
The approach to scaling seems to become smoother than that for the original 
Wilson action.  The organization of the paper is as follows: In section 2  we 
define the action we investigate and briefly recapitulate the definitions 
of various observables used and their scaling laws.  We present the detailed
results of our simulations in the next section and the last section  contains
a brief summary of our results and their discussion.

\bigskip
\begin{center}
\bf 2.  THE MODEL AND THE OBSERVABLES \\
\end{center}
\bigskip

\begin{figure}[htbp]\begin{center}
\epsfig{height=12cm,width=8cm,file=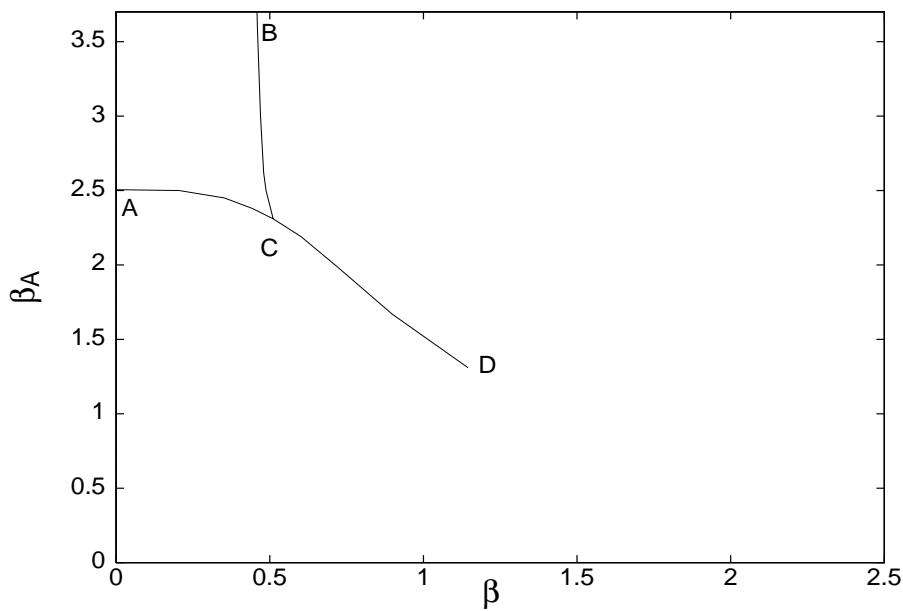,angle=270}
\caption{The phase diagram for the action (\protect\ref{sbc}),
showing the first order bulk phase transition lines. Taken from Ref.
\protect\cite{BhaCre} but with the endpoint {\bf D} as obtained in Ref.
\protect\cite{self}.} 
\label{fg.phase}\end{center}\end{figure}

Bhanot and Creutz\cite{BhaCre} found that the lattice theory defined by
the extended action of eq.(\ref{sbc}) has a rich phase structure, shown
in Fig. 1.  Similar results were obtained for the Villain action (\ref{sv}) by
Halliday and Schwimmer\cite{CaHaSc}.  In either case, the $\beta = 0$
axis describes an $SO(3)$ model which has a first order phase transition,
denoted by point {\bf A} in Fig. 1.  At $\beta_A ({\rm or}~\beta_V) =
\infty$, the theory reduces to a $Z_2$ lattice gauge theory with again a
first order phase transition at $\beta^{\rm crit} = {1\over2} \ln(1 +
\sqrt{2})$ $\approx$ 0.44 \cite {Weg}.  For both the mixed actions, these 
first order transitions extend into the coupling plane, as shown in
Fig. 1 by continuous lines. These lines meet at a triple point {\bf C} and
continue as a single line of first order transitions which ends at an
apparently critical point {\bf D}.  The proximity of {\bf D} to the
$\beta_A = 0$ line, which defines the Wilson action, has commonly been
held responsible for the abrupt change from the strong coupling region
to the scaling region for the Wilson action. It has also been attributed as a
possible reason for the dip in the non-perturbative $\beta$-function
obtained by Monte Carlo Renormalization Group methods.  Indeed, its
relative closeness to the $\beta_A = 0$ line for the $SU(2)$ theory
compared to the $SU(3)$ theory has been thought\cite{subet}  of as a
possible reason for the shallower dip in the former case.

The bulk transition in the Villain form of the SO(3) gauge theory is
known \cite{HalSch} to be caused by a condensation of $Z_2$ monopoles
in the strong coupling phase.  Defining $\sigma_c = \prod_{p \in
\partial c} \sigma_P$ for an elementary cube $c$, one finds that
$\sigma_c = -1$ characterizes a monopole located in $c$.  These
monopoles are absent in the weak coupling region, and can be suppressed
at stronger couplings by adding a term, $\lambda_M \sum_c \left( 1 -
\sigma_c \right)$ to the action (\ref{sv}) and setting $\lambda_M$
large.  Using this extra term, Ref. \cite{DatGav} demonstrated a clear
merging of the second order deconfinement line with a first order bulk
transition line for $\lambda_M =$ 1 for the Villain action (\ref{sv}).
Moreover, {\it two separate} transitions were shown to exist on the
{\it same} lattice near the merging point, thereby shedding some light
on the paradoxes a]-c] above. While pointing to the peculiar role the
bulk transitions play in affecting the deconfinement transitions, these
results also suggested that $Z_2$ electric current loops or vortices
have to be suppressed in restoring the universality for the mixed
action fully.   Defining $\sigma_l$ for a link $l \equiv (x, \hat \mu)$
at a point $x$ on the lattice in the $\mu$th direction as a product of
$\sigma_P$ of all those plaquettes which share the link $l$, i.e,
$\sigma_l = \prod_{p \in \hat \partial l} \sigma_P$, one finds that
$\sigma_l = -1$ signals the link to be a part of an $Z_2$ electric
current loop.  Adding further a term $\lambda_E \sum_l \left( 1 -
\sigma_l \right)$ to the action (\ref{sv}) in addition to the monopole
suppression term above, Refs. \cite{GavMan,DatGav} showed that
universality is restored in the entire coupling plane for large
$\lambda_E$.

While a similar mechanism is expected to work for the Bhanot-Creutz mixed
action (\ref{sbc}) as well, it is clear that both the monopole and 
vortex suppression terms added above do not depend on the gauge variables 
$U_\mu(x)$ directly and have to be generalized suitably for addition to
$S_{BC}$.  One possible way is to define $\sigma_c$ and $\sigma_l$ by
replacing $\sigma_P$ in them by sign(${\rm Tr}_F U_P$).  Thus the mixed
action with chemical potentials for these monopoles and vortices in that case 
is given by

\begin{eqnarray}
S_{BC,S}  =&&\sum_P \left( \beta \left( 1 - {1 \over 2} {\rm Tr}_F U_P
\right) + \beta_A \left( 1 - {1 \over 3} {\rm Tr}_A U_P \right) \right)
\nonumber \\ 
&&+ \lambda_M \sum_c \left(1 - \sigma_c \right) + \lambda_E \sum_l \left( 1 - 
\sigma_l \right),
\label{bcs}
\end{eqnarray}
where $\sigma_c = \prod_{p \in \partial c} {\rm sign (Tr}_F U_P)$ and
$\sigma_l = \prod_{p \in \hat \partial l} {\rm sign (Tr}_F U_P)$.
Comparing the naive classical continuum limit of eq. (\ref{bcs}) 
with the standard $SU(2)$ Yang-Mills action, one obtains
\begin{eqnarray}
{1 \over g^2_u} = {\beta \over 4} + {2\beta_A \over 3}~~.~~
\label{gu} 
\end{eqnarray}
Here $g_u$ is the bare coupling constant of the continuum theory.
Since the asymptotic scaling equation for the above mixed action with
additional (irrelevant) couplings $\lambda_M$ and $\lambda_E$ can be
easily written down in terms of $g_u$, it is clear that the
introduction of a non-zero $\beta_A$, $\lambda_M$, or $\lambda_E$ does
not affect the continuum limit: the theory for each $\beta_A$,
including the usual Wilson theory for $\beta_A = 0.0$, flows to the same
critical fixed point, $g^c_u = 0$, in the continuum limit for all
$\lambda_M$ and $\lambda_E$ and has the same scaling behavior near the
critical point.

One sees that even for $\beta_A = 0$ eq. (\ref{bcs}) corresponds to
a modified Wilson action with different densities of the monopoles and vortices
depending on the values of $\lambda_M$ and $\lambda_E$ respectively.  By
analogy with the works of Refs. \cite{GavMan,DatGav} for the Villain
action, one expects to eliminate all bulk transition lines in Fig. 1 by
setting $\lambda_M$ and $\lambda_E$ large.  In particular, the critical point
{\bf D} is expected to be absent in that case, leading perhaps to
a lot smoother transition from the strong coupling region to the scaling 
region for those values.  It has to be stressed though that universality 
has to be tested afresh for eq.(\ref{bcs}) even for $\beta_A = 0$ to be sure 
that the above naive argument about the $\lambda_M$ and $\lambda_E$ terms being 
irrelevant is correct. This is what we do in the following by determining a 
critical index of the deconfinement phase transition. We then check
whether the passage to scaling is affected by studying the deconfinement
transition as a function of the temporal lattice size.

It is interesting to note that no similar change occurs on the fundamental axis
($\beta_V = 0$) in the case of the Villain action since the variables
$\sigma_P$ are decoupled from the gauge variables in that case and can be
integrated out exactly for any observable depending solely on
$U_\mu(x)$.  However, even in that case the $\sigma_c$ and $\sigma_l$
could have been defined in terms of sign(${\rm Tr}_F U_P$) and similar
results as we obtain below would be obtained.
 
We studied the deconfinement phase transition on $N_\sigma^3 \times
N_\tau$ lattices by monitoring its order parameter and the
corresponding susceptibility for $N_\tau =$ 4, 5, 6 and 8 and $N_\sigma
=$ 8, 10, 12, 14, 15, and 16.  We used the simple Metropolis algorithm
and tuned it to have an acceptance rate $\sim 40$ \%.  The expectation
values of the observables were recorded every 20 iterations to reduce
the autocorrelations.  Errors were determined by correcting for the
autocorrelations and also by the jack knife method. The observables
monitored were the average plaquette, P, defined as the average of
${\rm Tr}_F U_P$/2 over all independent plaquettes, and the absolute
value of the average of the deconfinement order parameter \cite{McSv},
$L(\vec n)$, over the three dimensional lattice spanned by $\vec n$,
where $L$ is defined by
\begin{eqnarray}
L(\vec n) = {1\over2} {\rm Tr}_F \prod^{N_\tau}_{\tau=1} U_0 (\vec n,\tau)~.~
\label{pol}
\end{eqnarray}
Here $U_0 (\vec n,\tau)$ is the timelike link at the lattice site $(\vec
n,\tau)$.  In order to monitor the nature of
deconfinement and bulk phase transitions, we also define the
susceptibilities for both $|L|$ and P: 
\begin{eqnarray}
\chi_{|L|}~ =&&N_\sigma^3~~(\langle L^2 \rangle - \langle |L| \rangle ^2) \\
\label{chiL}
\chi_P~ =&&6 N_\sigma^3 N_\tau~~(\langle P^2 \rangle - \langle P \rangle^2)~.~
\label{chiB}
\end{eqnarray}

According to the finite size scaling theory\cite{Barb}, the peak of the
$|L|$ (or plaquette) susceptibility at the location of the deconfinement
(or bulk) transition should grow on $N_\sigma^3 \times N_\tau$ lattices
like 
\begin{equation} 
\chi^{\rm max}_{|L|~{\rm or}~P} \propto N_\sigma^{\omega}~,~ 
\label{chifs} 
\end{equation} 
for fixed $N_\tau$.
For a second order transition, $\omega \equiv \gamma/\nu$, where $\gamma$ and
$\nu$ characterize the growth of the $|L|$ (plaquette)-susceptibility
and the correlation length near the critical temperature (coupling)on an
infinite spatial lattice.  If the phase transition were to be of first
order instead, then one expects \cite{ChLaBi} the exponent $\omega = 3$,
corresponding to the dimensionality of the space.  In
addition, of course, the average $|L|$ or plaquette is expected to
exhibit a sharp, or even discontinuous, jump and the corresponding
probability distribution should show a double peak structure in case of
a first order phase transition.  Such an analysis of the
$|L|$-susceptibility for the Wilson action, where only $\beta$ is
nonzero in eq. (\ref{bcs}), yielded \cite{EngFin} an exponent $\omega =
1.93 \pm 0.03$, in good agreement with the corresponding  value ($1.965
\pm 0.005$) for the three dimensional Ising model, and the universality
conjecture\cite{SveYaf}.  Universality of the continuum limit of
lattice gauge theories predicts a similar deconfinement transition
belonging to the same universality class as the three dimensional Ising
model for all values of $\beta_A$, $\lambda_M$ and $\lambda_E$.

\bigskip 

\begin{center}
\bf 3. RESULTS OF THE SIMULATIONS\\
\end{center}
\bigskip
 
\begin{figure}[htbp]\begin{center}
\epsfig{height=12cm,width=9cm,file=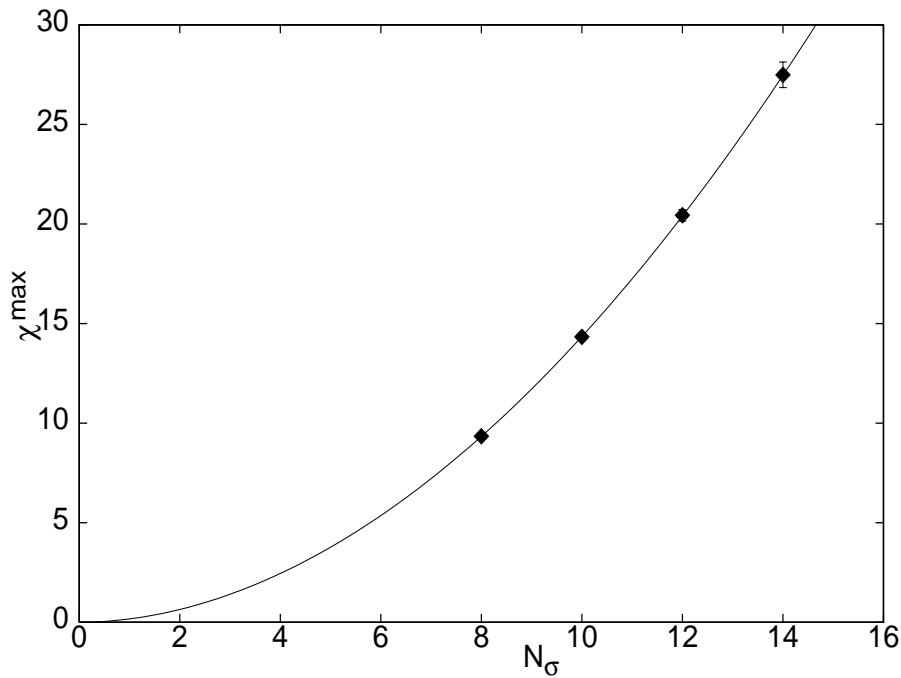,angle=270}
\caption{Variation of the peak height of $|L|$-susceptibility 
with spatial size $N_\sigma$ for fixed $N_\beta = 4$.
The curve is a fit to the eq.(\protect{\ref{chifs}}). }
\label{fg.chi4}\end{center}\end{figure}

In view of the fact that even for $\beta_A = 0$, the action (\ref{bcs})
differs from the Wilson action in a non-trivial manner for nonzero $\lambda_M$
and $\lambda_E$, we concentrate here on $\beta_A = 0$ and large $\lambda_M$ and
$\lambda_E$.  Our aim is to check the impact of the suppression of
monopoles and vortices, defined by the sign(${\rm Tr}_F U_P$) as above, on the
critical exponent $\omega$ on lattices with fixed $N_\tau $. We then wish
to study the scaling behaviour of the deconfinement transition by varying
$N_\tau$.  We chose $\lambda_M = 1$ and $\lambda_E = 5$ throughout this work.
Variations with respect to these as well as $\beta_A$ should in principle be
investigated although one would expect the results to display universality for
sufficiently large $\lambda_M$ and $\lambda_E$ if the $\beta_A = 0 $ results do so. 

\begin{center}
3.1  $N_\tau = 4 $\\
\end{center}
\bigskip
 
The deconfinement phase transition on $N_\sigma^3 \times 4 $ lattices for
$N_\sigma =$ 8, 10, 12 and 14 lattices was studied by first making
short hysteresis runs on the smallest lattice to look for abrupt or
sharp changes in both the average plaquette $\langle P \rangle$ and the
order parameter $ \langle |L| \rangle$. In the region of strong variation of
$\langle |L| \rangle$, longer runs were made to check whether the 
$|L|$-susceptibility exhibits a peak.  Histogramming
technique\cite{FerSwe} was used to extrapolate to nearby couplings for
doing this.  A fresh run was made at the $\chi_{|L|}$ peak position and the 
process repeated until the input coupling for the run was fairly close to the
output peak position of the susceptibility.  The same procedure was used
for the bigger lattices also  but by starting from the $\beta_c$ of the 
smaller lattice.  No peak was found in the vanishingly small plaquette 
susceptibility throughout, suggesting a lack of any nearby bulk transition.  
This should be contrasted\cite{spht} with the results for the 
$\lambda_M = \lambda_E = 0$, which is known to exhibit a peak.  Typically 
100-200 thousand measurements (2-4 million Monte Carlo iterations) were used
to estimate the magnitude of the peak height and the peak location for each
$N_\sigma$.   Table 1 lists our final results for all the $N_\sigma$ used.
The errors on $\beta_c$ were estimated by varying the bin size while
those for $\chi^{\rm max}$ were taken to be the maximum of the errors for all 
bin sizes.  Fitting the peak heights in Table 1 to eq.(\ref{chifs}), we 
obtained
\begin{equation}
\omega = 1.93 \pm 0.03 ~~.~~ \label{omval}
\end{equation}
Fig. \ref{fg.chi4} displays the very good quality of the fit.  The critical 
exponent $\omega$ is in excellent agreement with the values for both the 
standard Wilson action\cite{EngFin} and the 3-dimensional Ising model 
quoted in previous section.  Fitting the peak locations $\beta_{c,N_\sigma}$ 
by the usual 
finite size scaling expression,
\begin{equation}
\beta_{c,N_\sigma} = \beta_{c,\infty}  + B / N_\sigma^{1/\nu}~~,~~
\label{betafs}
\end{equation}
where $\nu$ = 0.63 is the correlation length exponent for the
3-dimensional Ising model and $B$ is a constant, we obtained 
$\beta_{c,\infty} = 1.326 \pm 0.006$, which is shifted by about one from 
the corresponding value for the $\lambda_M = \lambda_E = 0$ case which is 
2.2986 $\pm$ 0.0006 \cite{EngFin}.  Inspired by the agreement of $\omega$ 
above, we assumed universality to be true for $\nu$ as well in eq. 
(\ref{betafs}).  However, any reasonable variation of $\nu$ between 0.33 
and 1 changes the infinite volume extrapolation for $\beta_c$ by a few 
\begin{figure}[htbp]\begin{center}
\epsfig{height=14cm,width=10cm,file=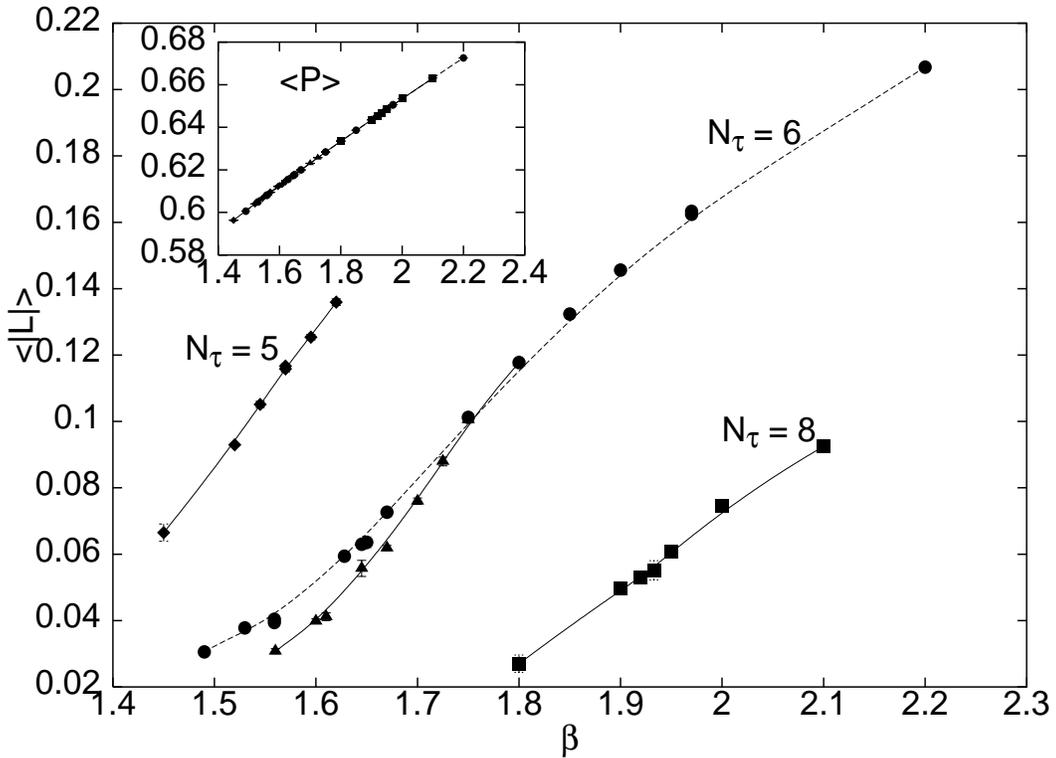,angle=270}
\caption{The deconfinement order parameter $\langle |L| \rangle$ as a function
of $\beta$ for various temporal lattices $N_\tau$.  For $N_\tau$ = 6, the 
circles (triangles) correspond to a $12^3$($15^3$) spatial lattice.  
The inset displays average plaquette $\langle P \rangle$ on all 
$N_\tau$ as a function of $\beta$. The lines are smooth extrapolations 
of the data to guide the eye.}
\label{fg.elp}\end{center}\end{figure}
($\sim 2$-3) per cent only. Thus the shift $\delta \beta = \beta_c^{
\rm Wilson} - \beta_c$ = 0.97 appears to be dominantly due to nonzero
$\lambda_M$ and $\lambda_E$, i.e, suppression of monopoles and
vortices.

\begin{center}
3.2 $N_\tau = 5 $, 6 and 8\\
\end{center}
\bigskip

In order to minimize finite spatial volume effects, we chose to work 
with $N_\sigma \ge 2 N_\tau$ always, as seen in sec. 3.1.  Consequently, the
full 4-volume $N_\sigma^3 N_\tau$ increased rapidly as we increased
$N_\tau$. This resulted in progressive shrinking of the coupling interval 
in which the histogramming technique was reliable.  We therefore
used many longer runs in the region of strong variation of $\langle |L|
\rangle$ to obtain the susceptibility directly and used the histogramming only
for the finer determination of the critical coupling.  Fig. \ref{fg.elp}
exhibits our results for $\langle |L| \rangle$ and $\langle P \rangle$
(shown in the inset) as a function of $\beta$ for $N_\tau$ = 5, 6 and 8.
A deconfinement phase transition is clearly visible for all of them. The
behaviour of the order parameter for two spatial volumes $12^3$ and $15^3$ 
for $N_\tau$ = 6 also supports the existence of a transition.  This is
more clearly seen in the corresponding $\chi_{|L|}$ determinations,
shown in Fig. \ref{fg.chi6}.  The plaquette, on the other hand, describes 
\begin{figure}[htbp]\begin{center}
\epsfig{height=14cm,width=10cm,file=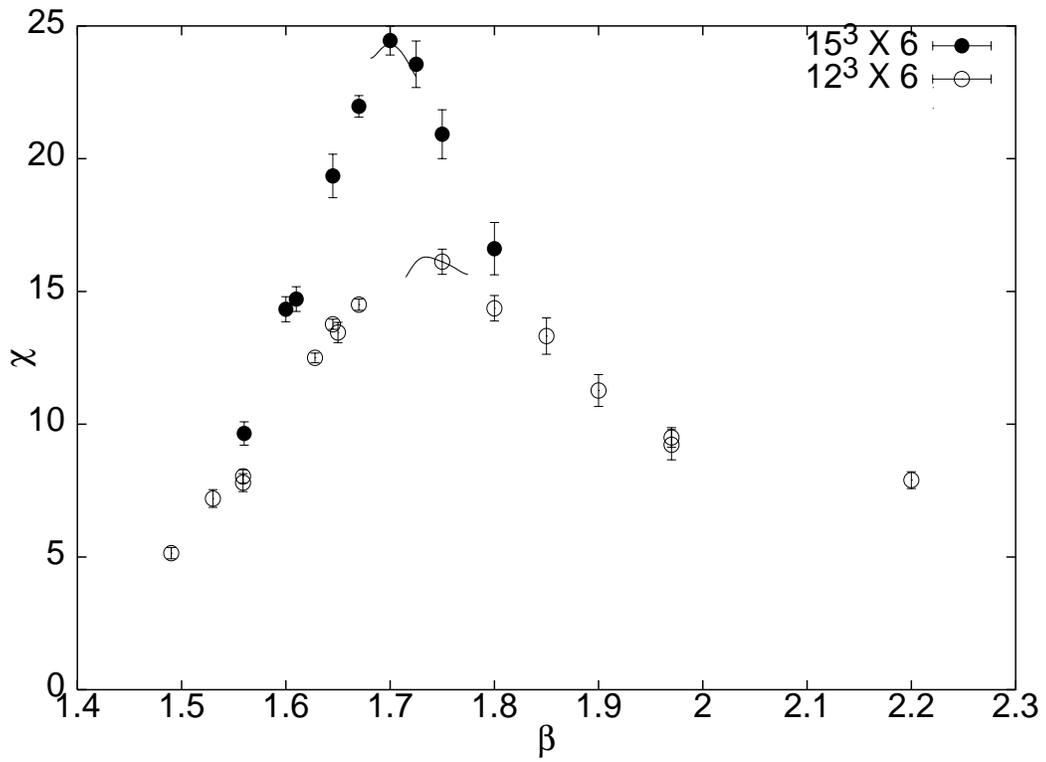,angle=270}
\caption{The $|L|$-susceptibility as a function of $\beta$ on lattices
with $N_\tau$ = 6. The continuous lines are extrapolations using the 
histogramming technique.}
\label{fg.chi6}\end{center}\end{figure}
a smooth and unique curve for all $N_\tau$ and $N_\sigma$ values. As the inset 
in Fig. \ref{fg.elp} shows, plaquette values for {\it all} these
lattices fall on the same curve, indicating an absence of any bulk
transition.  Tables 2 and 3 list the estimated maxima of $\chi_{|L|}$ 
for $N_\tau$ = 5 and 6 for two different spatial volumes along with the 
corresponding peak locations. As seen in Fig. \ref{fg.chi6} they are 
rather close to the input $\beta$ at which the long runs were made.  Using 
our value for $\omega$ from eq.(\ref{omval}), determined for $N_\tau$ = 4, 
and the peak height for the smaller spatial volume, the $\chi^{\rm max}$
on the bigger lattice can be predicted.  These predictions are listed in the 
respective tables in the last column and can be seen to be in very good 
agreement with the direct Monte Carlo determinations.  Alternatively, one 
can fit eq.(\ref{chifs}) to the peak heights in Tables 2 and 3 and 
determine $\omega$ again :
\begin{eqnarray}
\omega   =&& 1.99 \pm 0.17 ~~{\rm for}~~ N_\tau =5~~,~~ {\rm and} ~~
\nonumber \\ 
\omega   =&& 1.80 \pm 0.14 ~~{\rm for}~~ N_\tau =6~~.~~
\label{om56}
\end{eqnarray}
Both these determinations agree with the canonical values as well as our own
value in eq.(\ref{omval}). Not only is the universality of the
deconfinement phase transition thus verified on three different temporal
lattice sizes, but it also confirms that the same {\it physical} phase
transition is being simulated on them, thus approaching the continuum
limit of $a \to 0$ in a progressive manner by keeping the transition
temperature $T_c$ constant in physical units.

\begin{center}
3.3 Scaling of $T_c$ \\
\end{center}
\bigskip

\begin{figure}[htbp]\begin{center}
\epsfig{height=14cm,width=10cm,file=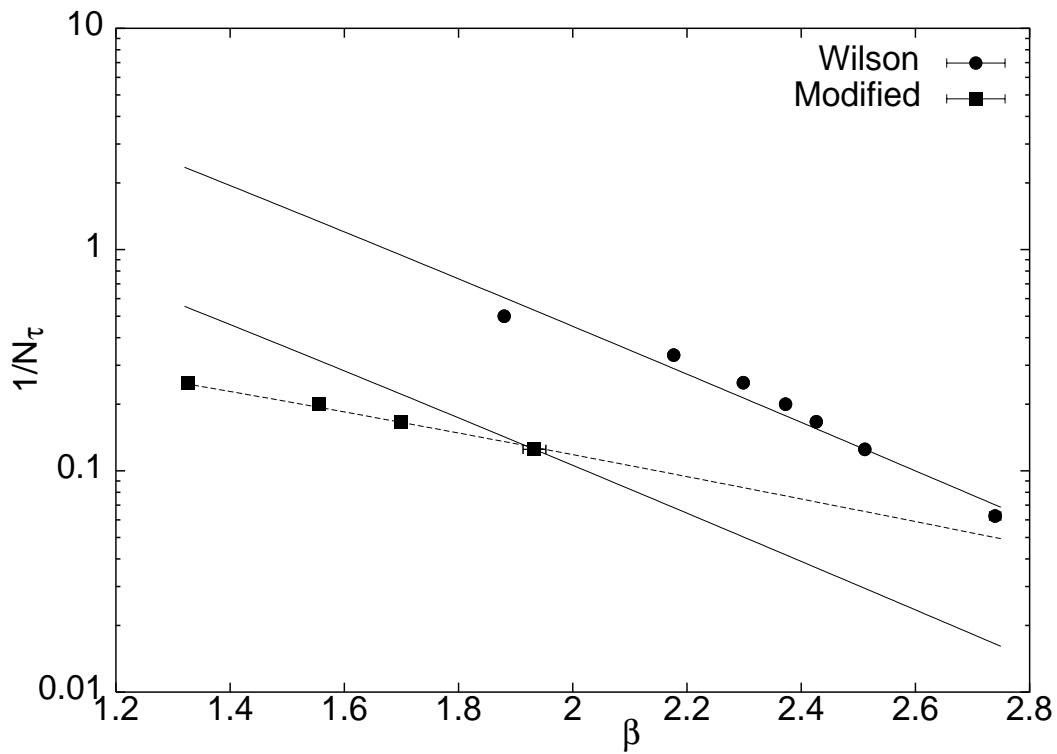,angle=270}
\caption{$1/N_\tau$ as a function of $\beta_c$. The squares are from this
work while circles are from Ref. \protect{\cite{Urs2}}.  The full lines depict 
the 2-loop asymptotic scaling relation of eq.(\protect{\ref{2lprg}}), 
normalized at $N_\tau$ = 8 in both cases.
The dashed line denotes eq.(\protect{\ref{rgs}}), normalized the same way. } 
\label{fg.scale}\end{center}\end{figure}

Table 4 lists the $\beta_c$ (estimated by extrapolating to infinite
volume whenever possible) for all the $N_\tau$ values we used. The
corresponding values for the usual Wilson action, i.e., $\lambda_M$ = 0 and
$\lambda_E$= 0 case, are also given in Table 4 along with the shifts caused by
switching on these two couplings. The shifts decrease with increasing
$N_\tau$ but nevertheless remains sizeable even for the largest lattice
we used. Their decrease smoothens the approach to the scaling limit, as we 
shall see below.  Fig. \ref{fg.scale} shows $aT_c = N_\tau^{-1}$
as a function of the corresponding critical $\beta$ for both our
simulations with suppression of monopoles and vortices and
the standard Wilson action (without any such suppression). The latter are
taken from the compilation of Ref. \cite{Urs2}. The full lines in the
figure show the 2-loop asymptotic scaling relation  
\begin{equation}
 aT_c = {1 \over N_\tau} \propto \left({4b_0 \over \beta }\right)^{-b_1/b_0^2} 
\exp \left(- {\beta \over 8b_0}\right)~~,~~
\label{2lprg}
\end{equation}
where 
\begin{equation}
b_0 = {11 \over 24 \pi^2}, ~~{\rm and}~~ b_1 = {17 \over 96 \pi^4}~~,~~
\label{b0b1}
\end{equation}
are the first two coefficients of the perturbative $\beta$-function for
the $SU(2)$ Yang-Mills theory.  The curve in each case was normalized 
to pass through the $N_\tau$= 8 data point.  The dashed line describes
a `phenomenological' scaling equation which is similar to the
eq.(\ref{2lprg}) but with the exponent increased by a factor of two:
\begin{equation}
 aT_c = {1 \over N_\tau} \propto \left({4b_0 \over \beta }\right)^{-b_1/b_0^2} 
\exp \left(- {\beta \over 4b_0}\right)~~.~~
\label{rgs}
\end{equation}

One sees deviations from asymptotic scaling for both the Wilson action and 
our action with suppression of monopoles and vortices.  The deviations
for the same range of $N_\tau$ seem larger for our action but then one
is also considerably deeper in the strong coupling region of the
Wilson action where one {\it a priori} would not have even expected any
scaling behaviour.  As the agreement of our results with the dashed line 
of eq.(\ref{rgs}) in Fig. \ref{fg.scale} shows, scaling may hold in 
this region of $\beta$ for the suppressed action, since 
the relation between $a$ and $\beta$ in this region (or $g^2$) is similar
to the asymptotic scaling relation, differing only in the exponent which 
will cancel in dimensionless ratios of physical quantities.
It is clear that as $\beta \to \infty$, the difference between
the two actions must vanish. The shifts in Table 4 do show such a trend
although the limiting point is not reached by $N_\tau$ = 8 definitely. 
It seems likely though that the trend of evenly spaced transition points
for our action will continue and the dashed line traced by its transition
points will merge with the Wilson action by $N_\tau \sim 25$ or so, as
suggested by its approach to the data for the
\begin{figure}[htbp]\begin{center}
\epsfig{height=14cm,width=10cm,file=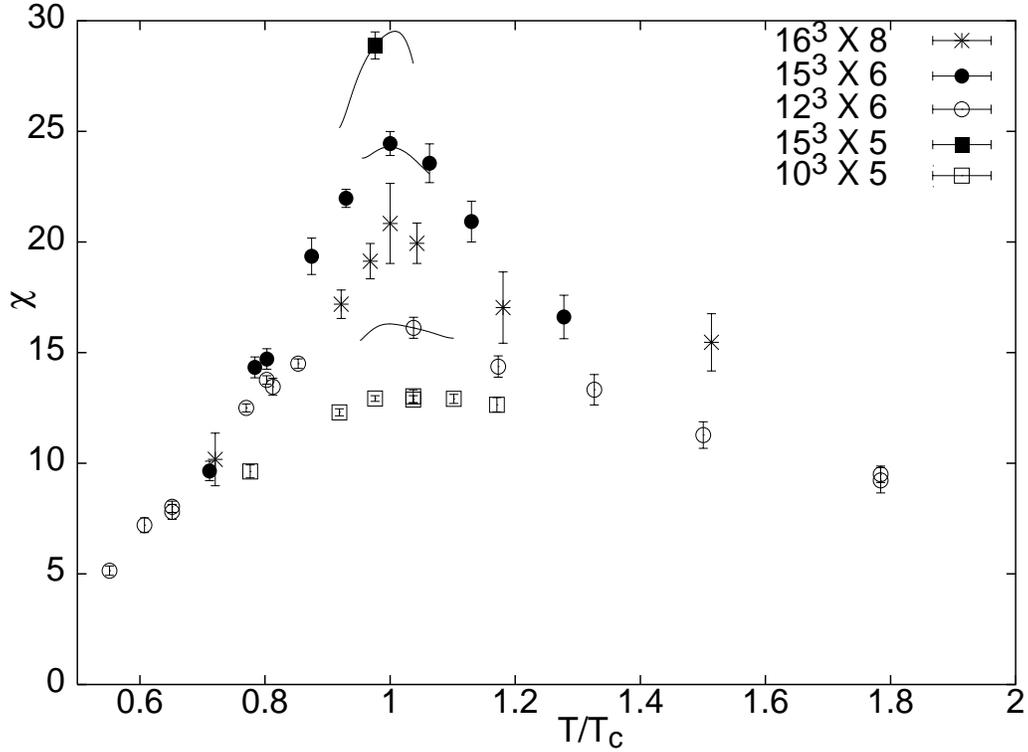,angle=270}
\caption{ Same as Fig. \protect{\ref{fg.chi6}} but as a function of $T/T_c$
and on additional lattices, as indicated.} 
\label{fg.allchi}\end{center}\end{figure}
Wilson action.  If this were to be so, a much smoother approach to
continuum limit is to be expected after the suppression of monopoles
and vortices. In particular, one expects that dimensionless ratios of
physical quantities at the deconfinement phase transition couplings
should be constant, already from $\beta \sim 1.33 $, which is the
transition point for the $N_\tau$ = 4.  

One possible interpretation of the results in Fig.  \ref{fg.scale} is
that the proximity of the point {\bf D} in Fig.  \ref{fg.phase} for the
usual Wilson action causes the nontrivial curvature visible in the data
for the Wilson action, and consequently its approach to the continuum
limit is not so smooth.  A strong suppression of monopoles and
vortices, as performed here, eliminates {\bf D}, resulting in a
smoother approach to scaling.  Of course, for very small lattice
spacings ( or large $\beta$), no significant difference between the two
will be seen but for sizeable values of the cut-off one may expect the
action with suppression to exhibit a better and smoother approach to
the continuum limit. We intend to check this by measuring the glueball
spectrum at the critical couplings for $N_\tau$ = 4--8.  In the
meantime, one can try to check this hypothesis by using eq.
(\ref{rgs}) to convert our $\chi (\beta)$ results to $\chi(T/T_c)$,
i.e, as a function of a dimensionless ratio for various $N_\tau$.
Fig.  \ref{fg.allchi} depicts the susceptibility $\chi$ as a function
of $T/T_c$ on lattices with $N_\tau$ = 5, 6 and 8.  Ideally one would
have expected all susceptibility data for the same physical volume to
fall on the same curve for different $N_\tau$.  These are the three
lowest curves with physical volume 8$T^{-3}$ but with $N_\tau$= 5, 6 and
8.  Unfortunately, the order parameter $\langle |L| \rangle$ is not
ultra-violet safe; it contains divergent contributions in the
continuum, making it $N_\tau$ (or $a$)-dependent even as a function of
$T/T_c$.  Consequently, the corresponding $\chi$'s are close but not on
any universal curve. On the other hand, an increase in physical volume
to 15.6 $T^{-3}$ (the $15^3 \times 6$ data) and 27$T^{-3}$ (the $15^3 \times
5 $ data) does seem to sharpen the susceptibility peak progressively,
as expected.  Although no quantitative analysis can be done
meaningfully due to the cut-off dependence of the order parameter
itself, the results do show the right trend and thus support a possible
scaling in the coupling region of these data points.


\begin{center}
4.  \bf SUMMARY AND DISCUSSION \\
\end{center}
\bigskip

The phase diagram of the mixed action of eq. (\ref{sbc}) in the
fundamental and adjoint couplings, $\beta$ and $\beta_A$, has been
a crucial input in understanding many properties of the $SU(2)$ and
$SU(3)$ lattice theories and their continuum limits.  The cross-over to 
the scaling region from the strong coupling region, as well as the dip 
in the non-perturbative $\beta$-function have been attributed to the 
location of the end point {\bf D} of the line of bulk first order phase 
transition.  In fact, even the relative shallowness of the dip for the 
$SU(2)$ case compared to the $SU(3)$ case is thought to be due to the 
closeness of the corresponding end point to the $\beta_A = 0$ Wilson
axis. Adding extra irrelevant terms to the action one obtains the
modified action of eq.(\ref{bcs}) in which monopoles and vortices can
be suppressed by setting the additional couplings to large values.
Based on the works\cite{GavMan,DatGav} for Villain action, one expects
the phase diagram to change completely in that case. In particular, no
phase transition lines or their critical end point {\bf D} will be
there, causing a smoother transition from the strong coupling region to
the scaling region.

In this paper we studied the deconfinement phase transition on the
fundamental axis in the $(\beta, \beta_A)$ coupling plane but with
$\lambda_M$ = 1 and $\lambda_E$ = 5, i.e., with strong suppression of
monopoles and vortices.  Our finite size scaling analysis yielded 1.93
$\pm$ 0.03 for the critical exponent $\omega \equiv \gamma/\nu$ for
lattices with $N_\tau$ = 4.  This value is in excellent agreement with
that\cite{EngFin} for the Wilson action and the three dimensional Ising
model, thus verifying the naive universality of the modified action.
However, as a result of the suppression, the critical coupling is
shifted by about unity compared to the Wilson case.  Our results on
$N_\tau$ = 5 and 6 also yielded similar values for $\omega$ albeit with
larger errors, confirming that the same physical transition was being
studied this way as a function of the lattice cut-off, $a$.  While the
$aT_c = N_\tau^{-1}$ was found to vary slower than expected from the
asymptotic scaling relation (\ref{2lprg}) for $N_\tau$ = 4--8, the data
did obey a similar relation with a factor of two larger exponent. A
straightforward extrapolation suggests the results from the modified
action will merge with those of Wilson action for large $N_\tau$ ( of
about $\sim$ 25), as expected in the limit of vanishing lattice spacing
$a$.  This suggests that the suppression makes the approach from the
strong coupling side to the scaling side much smoother than that for
the unsuppressed Wilson action, allowing us to simulate the theory at
smaller $\beta$.  It will be interesting to see if dimensionless ratios
of physical quantities such as glueball masses or string tension with
$T_c$ are constant in the range of critical couplings explored here.
Since the phase diagram for $SU(3)$, and indeed for $SU(N)$ lattice gauge
theories, is similar and the same mechanism is expected to work for
them, it will also be interesting to study such suppression in those
theories as well. However, additional possibilities for topological
objects may add further complications and may make it necessary to
suppress them as well.

\bigskip 

\begin{center}
6.  \bf ACKNOWLEDGMENTS \\
\end{center}
\bigskip 
It is a pleasure to acknowledge interesting discussions with Sourendu
Gupta.

\newpage 
\pagestyle{empty}
\begin{table}
\begin{center}
{Table 1~~~~~~~~~~~~~~~~~~~~}
\end{center}
The values of $\beta$ at which long simulations were performed \\
on $N_\sigma^3 \times 4$ lattices, $\beta_c$ and the height of 
the $|L|$-susceptibility \\ peak, $\chi^{\rm max}_{|L|}$.\\ \\
\medskip
\begin{tabular}{|c|c|c|c|}
\hline
                &           &               &                               \\
~~~~~$N_\sigma$~~~~~ &~~~~~ $\beta$ ~~~~~ & ~~~~~ $\beta_{c,N_\sigma}$ ~~~~~ & ~~~~~ $\chi_{|L|}^{\rm max}$~~~~~ \\
                &           &               &                               \\
\hline\hline 
                &           &               &                               \\
      8         &  1.37     &  1.366(7)     &     9.34 $\pm$ 0.07             \\
                &           &               &                               \\
\hline 
                &           &               &                               \\
     10         &  1.344    &  1.360(5)     &    14.34 $\pm$ 0.11             \\
                &           &               &                               \\
\hline
                &           &               &                               \\
     12         &  1.331    &  1.345(2)     &    20.44 $\pm$ 0.29             \\
                &           &               &                               \\
\hline
                &           &               &                               \\
     14         &  1.34     &  1.343(2)     &    27.48 $\pm$ 0.64             \\
                &           &               &                               \\
\hline\hline
\end{tabular} 
\end{table}

\begin{table}
\begin{center}
{Table 2~~~~~~~~~~~~~~~~~~~~}
\end{center}
Same as Table 1 but for on $N_\sigma^3 \times 5$ lattices.\\ \\
\medskip
\begin{tabular}{|c|c|c|c|c|}
\hline
&           &               &                &              \\
~~~~~$N_\sigma$~~~~~ &~~~~~ $\beta$ ~~~~~ & ~~~~~ $\beta_{c,N_\sigma}$ ~~~~~ & ~~~~~ $\chi_{|L|}^{\rm max}$~~~~~&~~~~~ $\chi^{\rm max}_{\rm predicted}$ \\
&           &               &                &              \\

\hline\hline 
&           &               &                &              \\
10         &  1.545    &  1.570(5)     &    13.17 $\pm$ 0.17&    --      \\
&           &               &                &              \\
\hline
&           &               &                &              \\
15         &  1.545    &  1.558(2)   &  29.57 $\pm$ 0.77& 28.78 $\pm$ 0.41 \\
&           &               &                &              \\
\hline\hline
\end{tabular} 
\end{table}

\begin{table}
\begin{center}
{Table 3~~~~~~~~~~~~~~~~~~~~}
\end{center}
Same as Table 1 but for on $N_\sigma^3 \times 6$ lattices.\\ \\
\medskip
\begin{tabular}{|c|c|c|c|c|}
\hline
&           &               &                &              \\
~~~~~$N_\sigma$~~~~~ &~~~~~ $\beta$ ~~~~~ & ~~~~~ $\beta_{c,N_\sigma}$ ~~~~~ & ~~~~~ $\chi_{|L|}^{\rm max}$~~~~~&~~~~~ $\chi^{\rm max}_{\rm predicted}$ \\
&           &               &                &              \\

\hline\hline 
&           &               &                &              \\
12         &  1.75     &  1.735(5)     & 16.34 $\pm$ 0.45 &   --      \\
&           &               &                &              \\
\hline
&           &               &                &              \\
15         &  1.70     &  1.702(2) &  24.39 $\pm$ 0.41&  25.13 $\pm$ 0.70  \\
&           &               &                &              \\
\hline\hline
\end{tabular} 
\end{table}

\begin{table}
\begin{center}
{Table 4~~~~~~~~~~~~~~~~~~~~}
\end{center}
The values of $\beta_c$ at which the deconfinement phase transition  \\
takes place on a lattice with temporal extension $N_\tau$  for eq.
(\ref{bcs}) \\ for $\lambda_M$ = 1 and $\lambda_E$ = 5 (column 2) and the usual
Wilson action \\ (column 3), taken from Ref.\cite{Urs2}. The last 
column lists the \\shift $\delta \beta = \beta_c^{\rm Wilson} - \beta_c$.\\
\medskip
\begin{tabular}{|c|c|c|c|}
\hline
                &           &            &               \\
~~~~~$N_\tau$~~~~~ &~~~~~ $\beta_c$ ~~~~~ & ~~~~~ $\beta_c^{\rm Wilson}$ ~~~~&~~~~$\delta \beta$~~~~\\
                &           &            &               \\
\hline\hline 
                &           &            &               \\
      4         &  1.327 $\pm$ 0.007     &  2.2986 $\pm$ 0.0006 &  0.972$\pm$0.007   \\
                &           &            &               \\
\hline 
                &           &            &               \\
      5         &  1.56 $\pm$ 0.01    &  2.3726 $\pm$ 0.0045 &   0.813$\pm$0.011  \\
                &           &            &               \\
\hline
                &           &            &               \\
      6         &  1.70 $\pm$ 0.01  &  2.4265 $\pm$ 0.0030 &  0.727$\pm$0.010   \\
                &           &            &               \\
\hline
                &           &            &   \\
      8         &  1.933 $\pm$ 0.02     &  2.5115 $\pm$ 0.0040 &  0.579$\pm$0.020   \\
                &           &            &   \\
\hline\hline
\end{tabular} 
\end{table}

\end{document}